# Title: Pascal conductance series in ballistic one-dimensional LaAlO$_3$/SrTiO$_3$ channels

**Authors:** Megan Briggeman[1,2], Michelle Tomczyk[1,2], Binbin Tian[1,2], Hyungwoo Lee[3], Jung-Woo Lee[3], Yuchi He[2,4], Anthony Tylan-Tyler[1,2], Mengchen Huang[1,2], Chang-Beom Eom[3], David Pekker[1,2], Roger S. K. Mong[1,2], Patrick Irvin[1,2], Jeremy Levy[1,2]

**Affiliations:** [1]Department of Physics and Astronomy, University of Pittsburgh, Pittsburgh, PA 15260, USA. [2]Pittsburgh Quantum Institute, Pittsburgh, PA 15260 USA. [3]Department of Materials Science and Engineering, University of Wisconsin-Madison, Madison, WI 53706, USA. [4]Department of Physics, Carnegie Mellon University, Pittsburgh, PA 15213, USA.

**One-sentence summary:** We report a sequence of quantized conductance plateaus that follows the third diagonal of Pascal's triangle.

**Abstract:** The ability to create and investigate composite fermionic phases opens new avenues for the investigation of strongly correlated quantum matter. We report the experimental observation of a series of quantized conductance steps within strongly interacting electron waveguides formed at the LaAlO$_3$/SrTiO$_3$ interface. The waveguide conductance follows a characteristic sequence within Pascal's triangle: $(1, 3, 6, 10, 15,...) \cdot e^2/h$, where $e$ is the electron charge and $h$ is the Planck constant. The robustness of these steps with respect to magnetic field and gate voltage indicate the formation of a new family of degenerate quantum liquids formed from bound states of $n = 2, 3, 4, ...$ electrons. These experiments could provide solid-state analogues for a wide range of composite fermionic phases ranging from neutron stars to solid-state materials to quark-gluon plasmas.

**Main Text:**
The investigation of strongly interacting fermionic systems and their resulting phases benefits from focusing on one-dimensional (1D) systems (*1-4*). By restricting the phase space for transport, correlations are significantly enhanced. A variety of techniques have been developed for understanding strongly-correlated 1D systems, ranging from Bethe ansatz to density matrix renormalization group (DMRG) approaches (*5*). Experimental investigations of degenerate, one-dimensional gases of paired fermions have been explored in ultracold atom systems with attractive interactions (*6*). In the solid state, attractive interactions have been engineered in carbon nanotubes using a proximal excitonic pairing mechanism (*7*). Electron pairing without superconductivity, indicating strong attractive interactions, has been reported in low-dimensional $SrTiO_3$ nanostructures (*8, 9*). However, color superfluids (*10*)—bound states of three or more particles—have only been observed in few-body bosonic systems (*11*).

$SrTiO_3$-based electron waveguides can provide insight into strongly interacting fermionic systems. The total conductance through an electron waveguide is determined by the number of extended subbands (indexed by orbital, spin, and valley degrees of freedom) available at a given chemical potential $\mu$ (*12, 13*). Each subband contributes one quantum of conductance $e^2/h$ with transmission probability $T(\mu)$ to the total conductance $G = (e^2/h)\sum_i T_i(\mu)$ (*14*). Quantized transport was first observed in III-V quantum point contacts (*15, 16*) and subsequently in 1D systems (*17-19*). Quantized conduction within 1D electron waveguides was recently demonstrated within $LaAlO_3/SrTiO_3$ heterostructures (*9*). A unique aspect of this $SrTiO_3$-based system is the existence of tunable electron-electron interactions (*20*) that lead to pairing (*8*) and superconductivity (*21*).

Here we investigate $LaAlO_3/SrTiO_3$-based 1D electron waveguides that are known to exhibit quantized ballistic transport as well as signatures of strong attractive electron-electron interactions and superconductivity. A thin film of $LaAlO_3$ (3.4 unit cells) is grown on $TiO_2$-terminated $SrTiO_3$ using pulsed laser deposition, using growth conditions that are described in detail elsewhere (*22*). Electrical contact is made to the interface in several locations that surround a given "canvas". An electron waveguide (Figure 1A) is created by first scanning a positively-biased ($V_{tip} \sim 10$ V) conductive atomic force microscope (c-AFM) tip in contact with the $LaAlO_3$ surface (*23, 24*). This process locally protonates (*25, 26*) the top $LaAlO_3$ surface and accumulates conducting electrons in the $SrTiO_3$ region near the $LaAlO_3/SrTiO_3$ interface. To restore an insulating state, negative voltages are applied to the tip, which locally de-protonates the $LaAlO_3$ surface. During a second pass of the tip along the channel, two negative voltage pulses ($V_{tip} \sim -15$ V) are used to create two weakly insulating, $L_B \sim 20-30$ nm-long barriers, separated by a length $L_S \sim 50-1000$ nm. A side gate is written ~1 μm away from the electron waveguide. Over a dozen specific devices have been investigated. Parameters and properties for seven representative devices A-G are given in Table 1.

The conductance of these electron waveguides depends principally on the chemical potential $\mu$ and the applied external magnetic field $\vec{B}$. The chemical potential is adjusted with a local side gate $V_{sg}$ (*9*), and for most experiments described here the external magnetic field is oriented perpendicular to the $LaAlO_3/SrTiO_3$ interface: $\vec{B} = B_z\hat{z}$. Quantum-point contacts formed in semiconductor heterostructures (*15, 16*) exhibit conductance steps which typically follow a linear sequence: $2 \times (1, 2, 3, 4, ...) \cdot e^2/h$, where the factor of 2 reflects the spin degeneracy. In an applied magnetic field, the electronic states are Zeeman-split, and resolve into steps of $(1, 2, 3, 4, ...) \cdot e^2/h$. By contrast, here we find that for certain values of magnetic field, the conductance steps for LaAlO$_3$/SrTiO$_3$ electron waveguides follow the sequence $(1, 3, 6, 10, ...) \cdot e^2/h$, or $G_n = n(n+1)/2 \cdot e^2/h$. As shown in Figure 1B, this sequence of numbers is proportional to the third diagonal of Pascal's triangle (Figure 1C, highlighted in red).

In order to better understand the origin of this sequence, it is helpful to examine the *transconductance $dG/d\mu$* and plot it as an intensity map as a function of $B$ and $\mu$. Transconductance maps for Devices A-F are plotted in Figure 2. A peak in the transconductance demarcates the chemical potential at which a new sub-band emerges; these chemical potentials occur at the minima of each subband, and as such we refer to them as subband bottoms (SBB). The peaks generally shift upward as the magnitude of the magnetic field is increased, sometimes bunching up and then again spreading apart. We observe many of the same features that were previously reported in 1D electron waveguides in LaAlO$_3$/SrTiO$_3$ (*27*) such as electron pairing and re-entrant pairing, which indicate the existence of electron-electron interactions. Near a special value of the magnetic field, locking of multiple subbands contribute to the total conductance as a function of chemical potential (see the labeled conductance plateaus in Figure 2A), which follows a Pascal series that is quantized in units of $e^2/h$.

Our approach to understanding the transport results described above begins with a single-particle description and incorporates interactions when the original description breaks down. Outside of the locked regions, the system is well described by a set of non-interacting channels, which places strong constraints on the theory of the locked regions. Any theory of the locked phases would need to both explain the locking of the transconductance peaks as well as quantized conductance steps away from the locked regime. Before settling on an attractive interaction interpretation of the locking phenomenon we considered a number of alternative mechanisms: spin-orbit, anharmonic confining potential, and impurity scattering but found that none of these resulted in the observed locking behavior. Within the single-particle description, we find that by fine-tuning the magnetic field and a single geometrical parameter of the waveguide, the ratio of vertical to lateral confinement strength, we can obtain the Pascal series of conductance plateaus. Next, we explore the addition of attractive electron-electron interactions to the model. The resulting calculations produce phases that are stable over a finite range of magnetic fields and geometrical parameters, thus lifting the requirement for fine-tuning that was imposed by the single-particle picture. In transconductance maps, the phases manifest themselves as the locking of peaks over a finite range of magnetic fields.

Our single-particle description excludes interactions but takes into account the geometry of the electron waveguide that produces the underlying subband structure. The four ingredients of the single-particle picture are electrons confined in the (1) vertical and (2) lateral directions by the waveguide and an external magnetic field that affects the electrons via the (3) Zeeman and the (4) orbital effect. *The intersection of more than two SBBs requires a special condition to be satisfied in the single-particle model.* The degeneracy requirement for obtaining the Pascal series (i.e. the crossing of 1, 2, 3, 4, ... SBBs) is satisfied by a pair of ladders of equispaced levels. Indeed, a pair of ladders of equispaced levels is naturally produced by a waveguide with harmonic confinement in both vertical and lateral directions. In the presence of Zeeman interactions, the waveguide Hamiltonian can be written (*9*)

$$H = \frac{(p_x - eB_z y)^2}{2m_x^*} + \frac{p_y^2}{2m_y^*} + \frac{p_z^2}{2m_z^*} + \frac{m_y^* \omega_y^2}{2} y^2 + \frac{m_z^* \omega_z^2}{2} z^2 - g\mu_B B_z s, \tag{1}$$

where $m_x^*$, $m_y^*$, and $m_z^*$ are the effective masses along the $x, y,$ and $z$ directions; $\omega_y$ and $\omega_z$ are frequencies associated with parabolic transverse confinement in the lateral ($y$) direction and half-parabolic confinement in the vertical ($z > 0$) direction, respectively; $g$ is the Landé factor; $\mu_B$ is the Bohr magneton; and $s = \pm 1/2$ is the spin quantum number. Eigenenergies corresponding to the SBBs are given by

$$E_{n_z, n_y, s} = \hbar\Omega \left(n_y + \frac{1}{2}\right) + \hbar\omega_z \left((2n_z + 1) + \frac{1}{2}\right) - g\mu_B B_z s, \tag{2}$$

where the electron eigenstates $|n_z, n_y, s\rangle$ are indexed by the orbital quantum numbers $n_z$ and $n_y$ and spin quantum number $s$, $\Omega = \sqrt{\omega_y^2 + \omega_c^2}$ is the magnetic field-dependent frequency associated with parabolic confinement of the electron in the lateral direction (being made of the bare frequency $\omega_y$ and the cyclotron frequency $\omega_c = eB_z/\sqrt{m_x^* m_y^*}$). To obtain two equispaced ladders of states we use the states associated with $\Omega$ for the first ladder and the states associated with $\omega_z$, split by the Zeeman splitting, for the second ladder. The Pascal series is produced by the "Pascal Condition": $\Omega = 4\omega_z = 2g\mu_B B_z/\hbar$. This condition requires fine-tuning of the magnetic field $B_z$ and the geometry of the waveguide ($\omega_y/\omega_z$). Meeting this condition results in crossings of increasing numbers of SBBs at a unique Pascal field $B_{Pa}$. By fitting the SBB energies given by Eq. (2) to experimental data, we are able to generate a peak structure (shown in Figure 3A) that is in general agreement with and has the same sequence of peak crossings as the experimentally observed transconductance. (Estimates for the single-particle model parameters are listed in Table 1.) By intentionally detuning the parameters away from the Pascal Condition (e.g., Figure 3B), the SBBs no longer intersect at a well-defined magnetic field. Fits of the single-particle model to experimental data for devices A-G (Figure 3C) show the expected correlation between $\omega_z$ and $\Omega(B_{Pa})$, but we do observe small deviations from the Pascal Condition for all samples.

The experimental data deviates from the single-particle model in several important ways. At low magnetic fields, the predicted linear Zeeman splitting of subbands is not obeyed; instead, the two lowest subbands ($|0,0,\pm 1/2\rangle$) are paired below a critical magnetic field, $B_P$ (9). At higher magnetic fields, re-entrant pairing is observed as subbands intersect and lock over a range of magnetic field values, near the Pascal field, $B_{Pa}$. In our non-interacting model (Eq. 1), there is a unique Pascal field $B_{Pa}$; however, experimentally we find that the value of the Pascal field depends on the degeneracy $n$: $B_{Pa}^{(n+1)} < B_{Pa}^{(n)}$. This shift of $B_{Pa}$ with the degeneracy may be due to an anharmonic component to the confinement. Adding an anharmonic term to the single particle model produces similar shifts of $B_{Pa}$ (see Section S7 for more details). The pairing field $B_P$ and Pascal field $B_{Pa}^{(2)}$ for devices A-G are shown in Table 1. Devices with similar geometries display a variety of pairing fields and Pascal fields. This is not unexpected based on previous work (8) where the pairing field was found to vary significantly from device to device and can be as large as $B_P = 11$ T. The cause for the differing strength of the pairing field is unknown but likely plays a role in the differing strengths of the locking for the Pascal degeneracies in this work. Fits of the transconductance data were made (shown in white on Figure 4A) for the $n = 2$ and $n = 3$ peak (or plateau) to determine if the states are, in fact, locking together over a finite range of magnetic fields. Details about the fitting procedure are described in Section S8 of the SOM. As shown in Figure 4B, the standard deviation of the fitted locations for the SBBs versus magnetic field for the $n = 2$ plateau, indicates a locking over a magnetic field range of $0.536 \pm 0.08$ T. This re-entrant pairing also occurs for the case of $n = 3$ intersecting modes (Figure 4C) where a fit of the transconductance data indicates that the states are locked together over a range of $1.14 \pm 0.15$ T. Re-entrant locking of as many as $n = 6$ ($G = 21\ e^2/h$) distinct intersections are observed (Figure 2, Device B). The Pascal series of conductance steps is observed for a variety of devices written with both short (50 nm) and long (1000 nm) electron waveguides, and at different angles, $\phi$, with respect to the crystallographic axis of the sample. These angles are listed in Table 1 and represent the angle of the waveguide device with respect to the (100) crystallographic direction. Devices with wires written at angles of 0°, 45°, or 90° show no significant difference.

A more sophisticated theoretical analysis is required to capture the effects of electron-electron interactions. In the absence of interactions, the single-particle model described by Equation (1) has band crossings but cannot predict any locking behavior. Prior work has demonstrated the existence of attractive electron-electron interactions in LaAlO$_3$/SrTiO$_3$ nanostructures (*8, 20*). We therefore construct an effective lattice model for the waveguide by extending the non-interacting model to include phenomenological, local, two-body interactions between electrons in different modes. This effective model is investigated using the density matrix renormalization group (DMRG), a numerical method which produces highly accurate results for strongly interacting systems in one dimension (*5, 28-33*). More details can be found in Section S2 of the online supplement.

The DMRG phase diagram in the vicinity of the $n = 2$ plateau is shown in Figure 5A. Distinct phases are illustrated by regions of solid color and identified with text in each representative region. In addition to a vacuum phase (V; no electrons) and a one-Fermi-sea phase (F; one mode is occupied), we also find a two-Fermi sea phase (2F; both modes occupied) and an electron-pair phase (P) in which single electron excitations are gapped out and both modes have equal electron density. Boundaries between phases are of two types: phase boundaries in which the number of fermion modes is unchanged (indicated by dashed lines) and phase boundaries in which the number of fermion modes changes (indicated by solid white lines). The solid white line boundaries correspond to peaks in the transconductance, which are also highlighted in the experimental data in Figure 4A. Experimentally, DC measurements cannot distinguish between phases that have the same conductance (e.g., P and 2F). We expect the pairing strength to scale as $U^2/t$, where $U$ is the attractive interaction strength and $t = \hbar^2/ma^2$ is the bandwidth (where $m$ is the band mass and $a$ is the lattice spacing). This energy scale determines the range of magnetic field over which electrons are locked together.

Extending this calculation to three electron modes with attractive interactions maps out the $n = 3$ plateau and its associated phases (Figure 5B). The phases associated with the $n = 3$ plateau are "trions", bound states of three electrons (T) that form a one-dimensional degenerate quantum liquid. In this phase, all one- and two-electron excitations are gapped out, but three-electron excitations are gapless. Adjacent to the trion phase are two related 3-electron phases: one in which a single electron breaks free, leaving behind a pair (F+P), and another in which all three fermions are independent (3F). The three phases are distinguished by the number of gapless modes but all contribute three conductance quanta to the DC conductivity. Other phases exist at lower chemical potentials (P, 2F, F, V), rounding out the entire phase diagram. We explore the stability of the phase diagram as we deviate from the Pascal Condition and find that the trion phase is remarkably stable, as compared to the competing phases (see section S6 of the online supplement), consistent with the findings in Figure 3C, which shows that the Pascal condition is not precisely met.

Phases of composite particles, like trions, are known to be fragile in higher dimensions, requiring strong attractive interactions (*10*). By contrast, previous investigations (*34, 35*), as well as our numerical simulations, indicate that these $n$-electron phases are very robust in one dimension. For example, the trion phase remains stable even if one of the three electron-electron interactions becomes repulsive (*36*). Consequently, we expect that transitions between the vacuum and $n$-electron phases (like the vacuum-trion transition), associated with jumps of the conductance by $(2, 3, 4, ...) \cdot e^2/h$, should be a prominent feature of multi-component one-dimensional systems. Wherever $n$ lines cross, attractive interactions should stabilize an $n$-electron composite phase.

Here we discuss other theoretical explanations that we have considered. The addition of spin-orbit coupling to the non-interacting model modifies the subband structure, producing avoided crossings of the transconductance peaks. Anharmonicity of the confining potential, in the absence of interactions, bends the subband structure but also does not produce locking. We rule out impurity scattering effects due to the ballistic nature of the transport. Moreover, without inter-electron interactions (e.g. negative U center (*37*)), an impurity cannot produce locking phenomena. We are not aware of other mechanisms for locking, but cannot rule them out. Finally, we remark, that any theory of the locking phenomenon would need to have a non-interacting limit that matches with experiments, e.g. predict conductance quantization.

The Pascal Condition assumes that the magnetic field is oriented out of plane. To investigate the effect of in-plane magnetic field components on the Pascal conductance series, we measure angle-dependent magnetotransport, with the magnetic field oriented at an angle $\theta$ with respect to the sample normal, within the $y - z$ plane, $\vec{B} = B(\sin \theta \; \hat{y} + \cos \theta \; \hat{z})$ (Figure 6A). In the out-of-plane orientation ($\theta = 0°$), characteristic Pascal behavior is observed, with subband locking taking place near 6 T (Figure 6D, $\theta = 0°$). As $\theta$ increases, the trion phase associated with the $n = 3$ plateau destabilizes, while another (non-Pascal series) trion phase forms in a different region of parameter space (Figure 6D, $\theta = 20°$, indicated by white lines). At larger angles (Figure 6D, $\theta = 50°$), a dense network of reentrant pairing, disbanding, and re-pairing is observed. (See SOM for a video showing the evolution of the transconductance spectra from $\theta = 0°$ to $\theta = 90°$.) The strength of the re-entrant pairing of the $|0,0,\downarrow\rangle$ and $|0,1,\uparrow\rangle$ subbands is strongly-dependent on the angle $\theta$ of the applied magnetic field (Figure 6C). The lower ($B_R^-$) and upper ($B_R^+$) magnetic fields over which these SBBs are locked together is indicated in Figure 6D with red and blue circles. The magnetic field range ($\Delta B_R = B_R^+ - B_R^-$) is shown as a function of angle (Figure 6C). The strength of the re-entrant pairing, $\Delta B_R$, initially increases with angle, jumps discontinuously at $\theta = 30°$, as the SBBs (which have been shifting closer) snap together, and then decreases again. At $\theta = 0°$ there is a non-Pascal series crossing (no locking) of like-spin states ($|0,0,\downarrow\rangle$, $|0,1,\downarrow\rangle$), highlighted by crossed lines, which evolves into an avoided crossing at $\theta = 10°$. This feature is explored in Figure 6B where we plot conductance curves at $B = 3$ T for different angles.

The "Pascal-liquid" phases reported here may constitute a new class of quantum degenerate electronic matter. Pascal composite particles would have a charge $ne$, where $n = 2, 3, 4, ...$, and as-yet-undetermined spin quantum numbers. As with fractional fermionic states, it seems likely that the expected charge could be verified from a shot-noise experiment (*38*). The particular Pascal sequence observed here experimentally is a consequence of the number of spatial dimensions in which they exist. Hypothetically, a material with four dimensions (three transverse to a conducting channel) could exhibit a conductance sequence $(1, 4, 10, 20, ...) \cdot e^2/h$, the next diagonal in the Pascal triangle. The Pascal sequence of bound fermions is reminiscent of the "quantum dot periodic table" used to categorize multi-electron states in semiconductor nanostructures (*39*); the difference here is that the Pascal liquids are comprised of composite particles that are free to move in one spatial dimension, held together by mutual attraction rather than by an external potential profile. Pascal composite particles with $n > 2$ can be regarded as a generalization of Cooper pair formation, analogous to the manner in which quarks combine to form baryonic and other forms of strongly interacting, degenerate quantum matter. Interactions among Pascal particles are in principle possible—for example, trions could in principle "pair" to form bosonic hexamers. Coupled arrays of 1D waveguides can be used to build 2D structures. This type of structure is predicted to show a wide variety of properties ranging from sliding phases (*40-42*) to non-abelian excitations (*43*). This highly flexible oxide nanoelectronics platform is poised to synthesize and investigate these new forms of quantum matter.

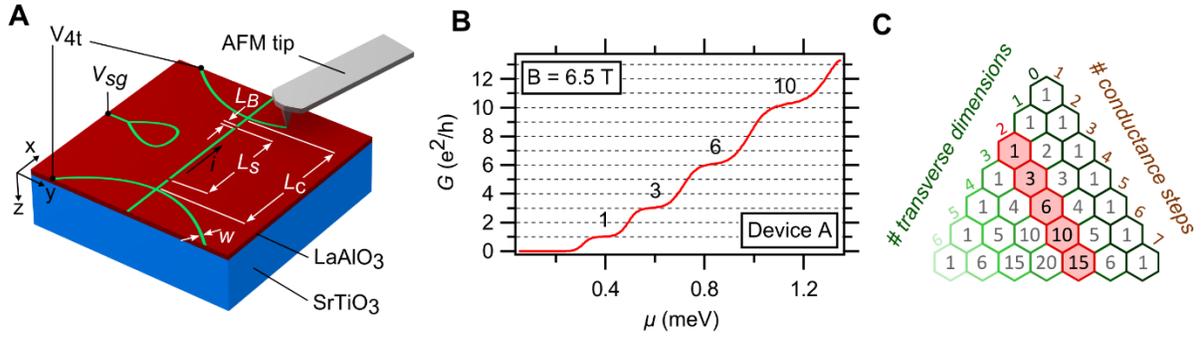

**Figure 1 Pascal series of conduction steps in an electron waveguide.** (**A**) Depiction of the sketched waveguide. Green lines indicate conductive paths at the LaAlO$_3$/SrTiO$_3$ interface. Device dimensions are indicated: barrier width $L_B$, barrier separation $L_S$, total length of the channel between the voltage sensing leads $L_C$, and nanowire width as measured at room temperature typically $w \sim 10$ nm. (**B**) Conductance $G$ through Device A at $T = 50$ mK and $B = 6.5$ T. A series of quantized conductance steps appears at $(1, 3, 6, 10, ...) \cdot e^2/h$. (**C**) Pascal triangle representation of observed conductance steps, represented in units of $e^2/h$. The highlighted diagonal represents the sequence for an electron waveguide with two transverse degrees of freedom.
8

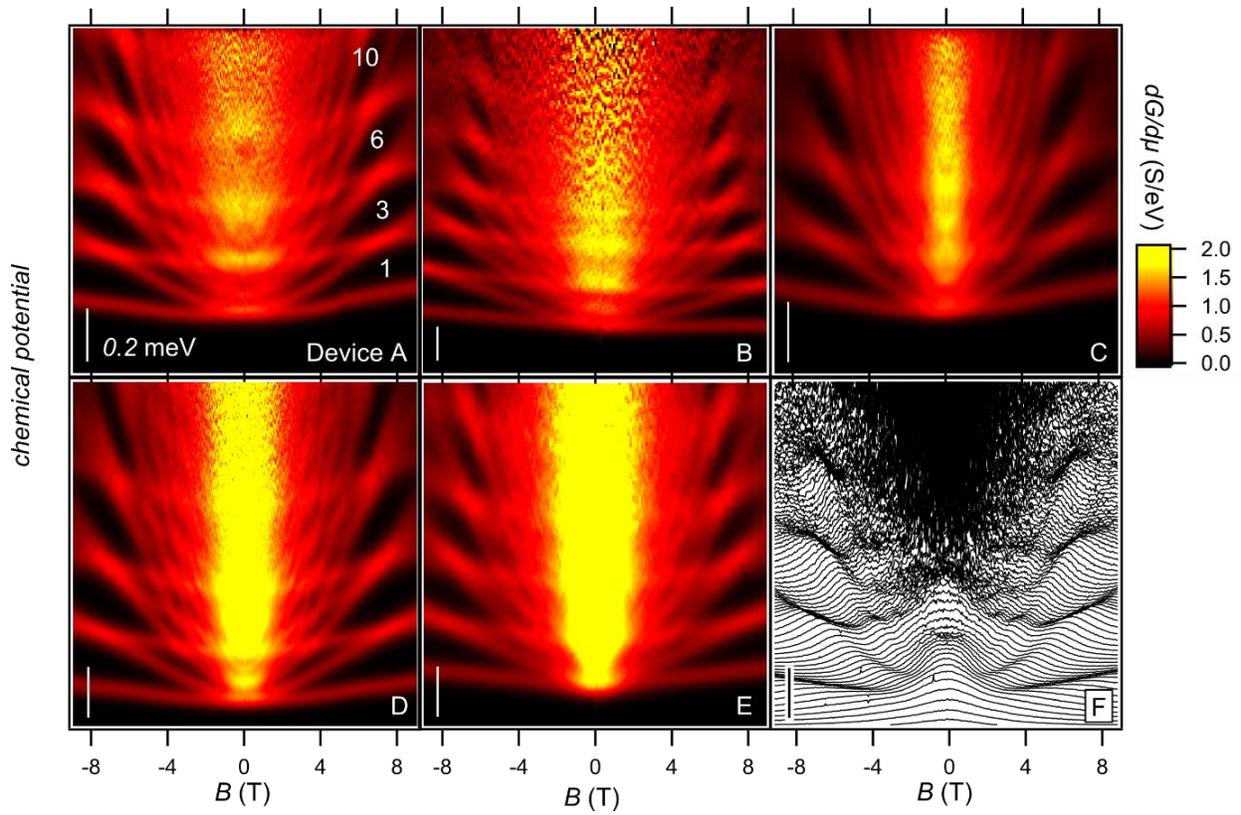

**Figure 2 Transconductance maps of Pascal devices**. Transconductance $dG/d\mu$ plotted as a function of chemical potential $\mu$ and out-of-plane magnetic field $B$ for representative devices A-F. Bright regions indicate increasing conductance as new subbands become occupied, and dark regions indicate conductance plateaus. Conductance values for several plateaus are indicated in white in panel A highlighting the Pascal series seen in all six devices shown here. Vertical scale bars in each panel represent 0.2 meV in chemical potential. The transconductance of Device F is displayed as a waterfall plot with vertical offsets given by the chemical potential at which the curve was acquired.

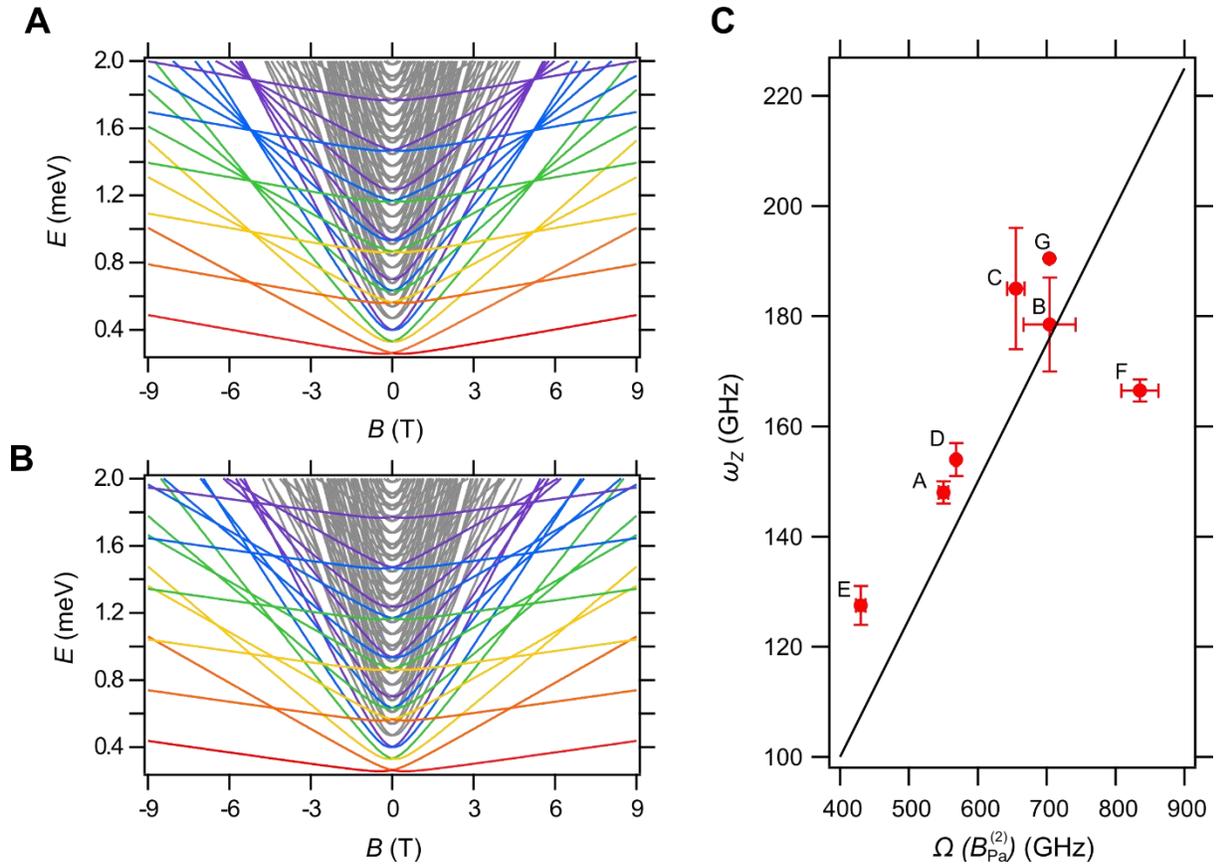

**Figure 3 Subband energies for non-interacting electron waveguide model.** (**A**) Energy $E$ vs $B$ calculated from the single-particle model, with parameters tuned to give Pascal degeneracies: $l_y = 33$ nm, $l_z = 10$ nm, $m_y = 1\,m_e$, $m_z = 5\,m_e$, $g = 1.0$. States are colored to highlight the bunching of increasing numbers of states to form the Pascal series conductance steps. (**B**) $E$ vs $B$ calculated from the single particle model, where the parameters have been detuned by 20%, destroying the crossing of electron states: $l_y = 33$ nm, $l_z = 10$ nm, $m_y = 1\,m_e$, $m_z = 5\,m_e$, $g = 1.2$. (**C**) Plot of $\omega_z$ vs $\Omega(B_{Pa}^{(2)})$ for Devices A-G, shows that while $\omega_z$ and $\Omega(B_{Pa}^{(2)})$ vary significantly from sample to sample they are all near the theoretically-predicted critical value of $\frac{\omega_z}{\Omega(B_{Pa})} = 0.25$, denoted by the solid black line.

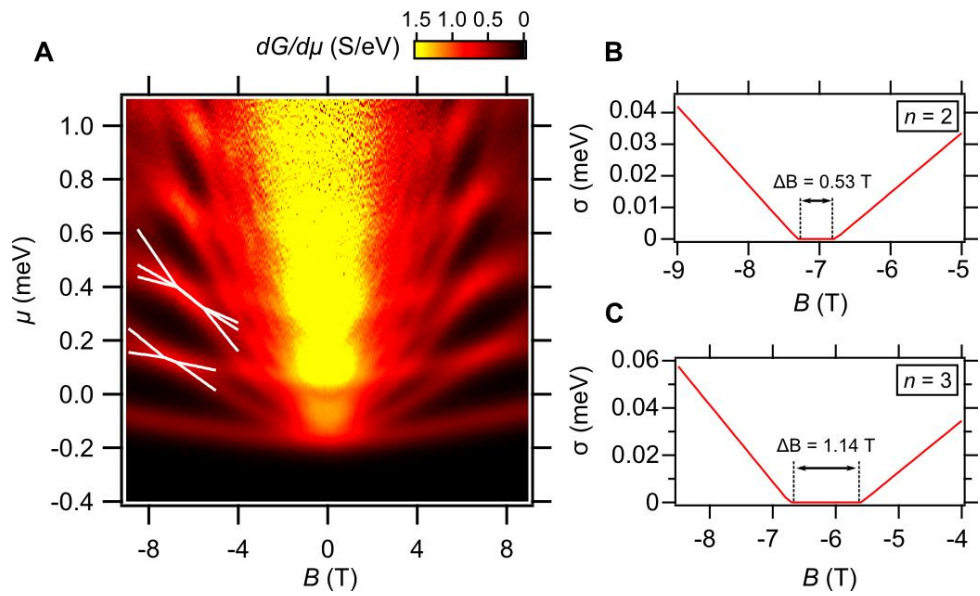

**Figure 4 Fits of Pascal conductance plateaus.** (**A**) Transconductance data for Device F. White lines are fits of the peak locations for the $n = 2$ and $n = 3$ Pascal plateaus and correspond to contribution of additional subbands in the transconductance data. (**B**) Standard deviation between states $|0,0,\uparrow\rangle$ and $|0,1,\downarrow\rangle$ from a fit of the experimental data in panel (A) showing the reentrant pairing as the states come together and are locked for a range of magnetic field values, indicated by $\Delta B$. (**C**) Fits for the $n = 3$ state also produce a standard deviation that shows the three states converge and are locked together for a range of magnetic field values, $\Delta B$. A description of the fitting method can be found in section S8 of the SOM.

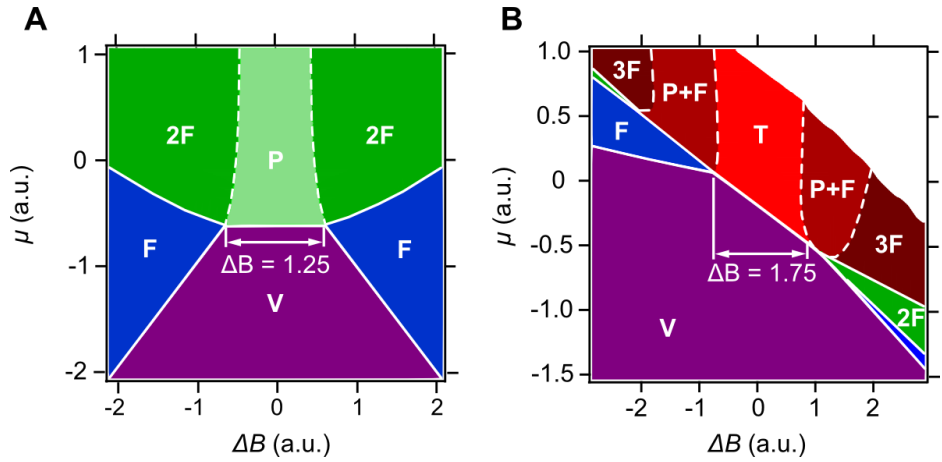

**Figure 5 DMRG phase diagrams.** DMRG phase diagrams calculated for two (**A**) and three (**B**) fermions with attractive interactions in one dimension. Abbreviations for various phases: $m$F: $m$ distinct fermi surfaces, P: paired phase, T: trion phase, V: vacuum, A+B: phase composed of A and B. Solid white lines correspond to the fits highlighted in Figure 4A. Details about the calculations can be found in the SOM. The black numbers on the plots indicate the strength of the locking for the pair (A) and trion (B) phases. Similar to what is observed in fits of the experimental data, the trion phase is locked over a larger range of magnetic field values.

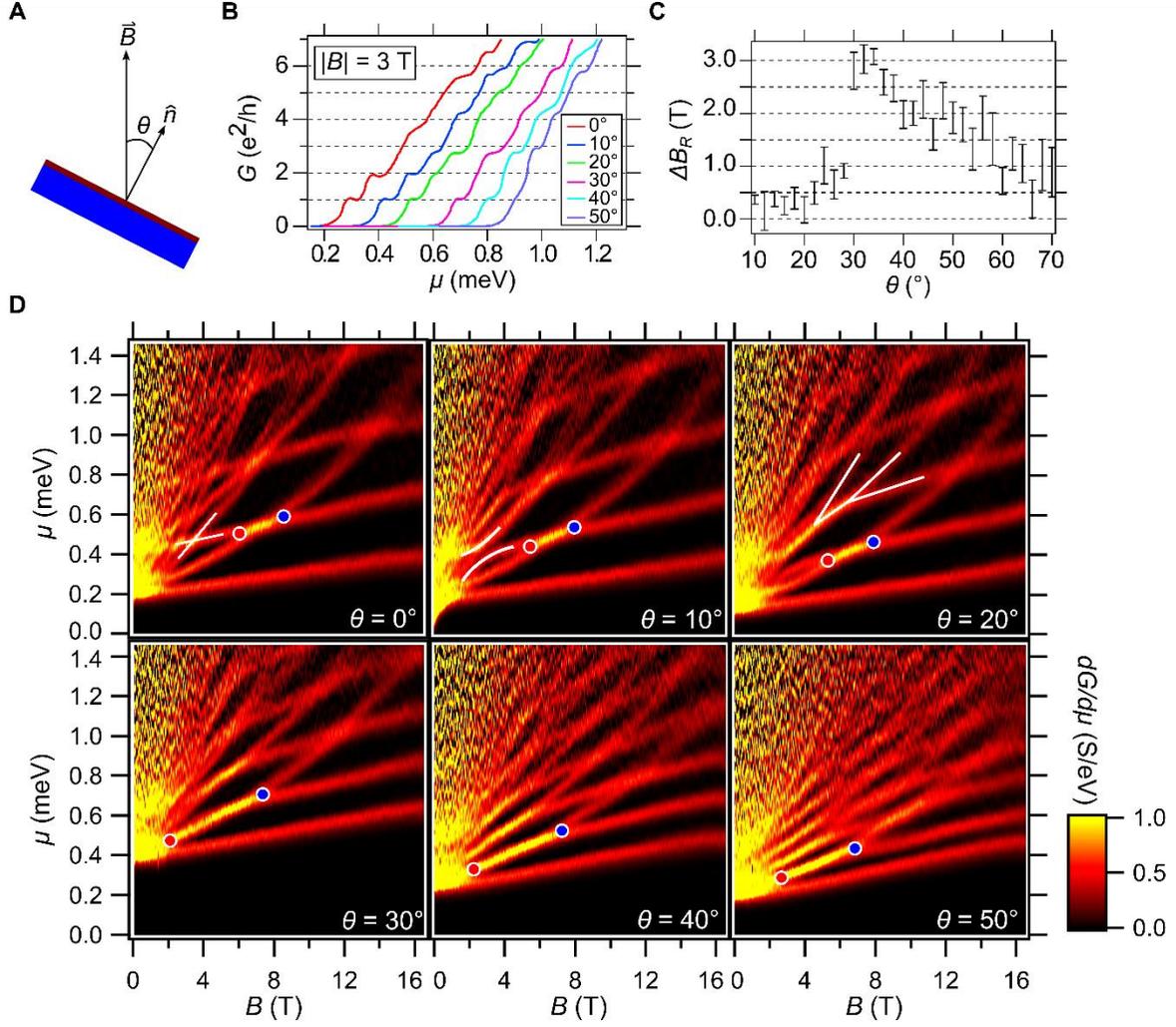

**Figure 6 Angle dependence of waveguide transport.** Waveguide device G. **(A)** Schematic of the sample as it is rotated with respect to the direction of the magnetic field $\vec{B}$. $\hat{n}$ is the vector normal to the plane of the sample and $\theta = 0°$ represents an out-of-plane magnetic field orientation. **(B)** Conductance curves as a function of angle at $|B| = 3$ T. As the magnetic field is rotated away from an out-of-plane angle, we see an avoided crossing open up, which can be seen in the $\theta = 10°$ curve as the plateau that begins to form near 3 $e^2/h$. We can also see evidence of the reentrant pairing that starts to occur at larger angles ($\theta > 30°$) when the conductance increases by a step of 2 $e^2/h$, from 1 $e^2/h$ to 3 $e^2/h$. **(C)** Reentrant pairing strength as a function of angle $\theta$. **(D)** Transconductance $dG/d\mu$ as a function of magnetic field strength and chemical potential. The magnetic field is rotated from an out-of-plane orientation ($\theta = 0°$) to $\theta = 50°$ in 10 degree steps. The in-plane component of the magnetic field is roughly perpendicular the waveguide channel. At small angles the Pascal series can be seen in the transport with bunches of 1, 2, and 3 sub-bands, but this is broken as the angle is increased. The reentrant pairing strength is indicated by the points where the states first lock together (red circles) and break apart (blue circles).

Table 1: **Parameters for waveguide devices A-G**. $L_B$ is the width of the barriers in the waveguide, $L_S$ is the separation between the barriers, $L_C$ is the total length of the channel between the voltage sensing leads, $\phi$ is the angle between the nanowire and the (100) crystallographic direction. $B_P$ indicates the pairing field, $B_{Pa}^{(2)}$ indicates the field at the center of the $n=2$ Pascal phase. Single particle model fits of experimental data determine the Landé factor $g$, the effective mass in the y direction $m_y$, frequencies associated with parabolic confinement in the lateral (y) direction and half-parabolic confinement in the vertical ($z > 0$) direction $\omega_y$ and $\omega_z$, and the effective trapping frequency in the y-direction $\Omega(B_{Pa}^{(2)})$.

| Device | $L_B$ (nm) | $L_S$ (nm) | $L_C$ (nm) | $\varphi$ (°) | $B_P$ (T) | $B_{Pa}^{(2)}$ (T) | $g$ | $m_y$ ($m_e$) | $\omega_y$ (GHz) | $\omega_z$ (GHz) | $\Omega(B_{Pa}^{(2)})$ (GHz) |
|---|---|---|---|---|---|---|---|---|---|---|---|
| A | 20 | 50 | 500 | 0 | 0.89 | 7.23 | 0.53 | 2.34 | 81 | 148 | 550 |
| B | 20 | 50 | 500 | 0 | 3.01 | 6.66 | 1.11 | 1.68 | 94 | 178 | 704 |
| C | 30 | 1000 | 1800 | 0 | 0.70 | 7.70 | 0.60 | 2.15 | 178 | 185 | 655 |
| D | 20 | 50 | 500 | 45 | 1.12 | 7.37 | 0.56 | 2.32 | 103 | 154 | 568 |
| E | 20 | 50 | 500 | 45 | 1.42 | 6.69 | 0.55 | 2.81 | 97 | 127 | 429 |
| F | 20 | 50 | 500 | 90 | 1.68 | 5.93 | 0.89 | 1.25 | 34 | 166 | 835 |
| G | 30 | 1000 | 1800 | 0 | 0.22 | 7.68 | 0.58 | 1.94 | 99 | 190 | 703 |

**List of Supplemental Materials:**
Supporting information Sections S1-S12
Fig S1-S12
References (44-47)
Movie S1: https://youtu.be/yBz3HaNi4jE

**References and Notes:**

1.  S. Tomonaga, Remarks on Bloch's Method of Sound Waves applied to Many-Fermion Problems. *Prog Theor Phys* **5**, 544-569 (1950).
2.  J. M. Luttinger, An Exactly Soluble Model of a Many-Fermion System. *Journal of Mathematical Physics* **4**, 1154-1162 (1963).
3.  T. Giamarchi, *Quantum Physics in One Dimension*. (Oxford University Press, 2003).
4.  D. C. Mattis, E. H. Lieb, Exact Solution of a Many‐Fermion System and Its Associated Boson Field. *Journal of Mathematical Physics* **6**, 304-312 (1965).
5.  S. R. White, Density matrix formulation for quantum renormalization groups. *Phys Rev Lett* **69**, 2863-2866 (1992).
6.  Y.-a. Liao *et al.*, Spin-imbalance in a one-dimensional Fermi gas. *Nature* **467**, 567 (2010).
7.  A. Hamo *et al.*, Electron attraction mediated by Coulomb repulsion. *Nature* **535**, 395-400 (2016).
8.  G. Cheng *et al.*, Electron pairing without superconductivity. *Nature* **521**, 196 (2015).
9.  A. Annadi *et al.*, Quantized Ballistic Transport of Electrons and Electron Pairs in $LaAlO_3/SrTiO_3$ Nanowires. *arXiv:1611.05127*, (2016).
10. Á. Rapp, G. Zaránd, C. Honerkamp, W. Hofstetter, Color Superfluidity and ``Baryon'' Formation in Ultracold Fermions. *Phys Rev Lett* **98**, 160405 (2007).
11. T. Kraemer *et al.*, Evidence for Efimov quantum states in an ultracold gas of caesium atoms. *Nature* **440**, 315 (2006).
12. R. Landauer, Spatial Variation of Currents and Fields Due to Localized Scatterers in Metallic Conduction. *Ibm J Res Dev* **1**, 223-231 (1957).
13. D. L. Maslov, M. Stone, Landauer conductance of Luttinger liquids with leads. *Phys Rev B* **52**, R5539-R5542 (1995).
14. S. Datta, *Quantum transport : atom to transistor*. (Cambridge University Press, Cambridge, UK ; New York, ed. Repr. with corrections., 2013), pp. xiv, 404 pages.
15. B. J. van Wees *et al.*, Quantized conductance of point contacts in a two-dimensional electron gas. *Phys Rev Lett* **60**, 848-850 (1988).
16. D. A. Wharam *et al.*, One-Dimensional Transport and the Quantization of the Ballistic Resistance. *J Phys C Solid State* **21**, L209-L214 (1988).
17. A. Yacoby *et al.*, Nonuniversal conductance quantization in Quantum wires. *Phys Rev Lett* **77**, 4612-4615 (1996).
18. S. Frank, P. Poncharal, Z. L. Wang, W. A. d. Heer, Carbon Nanotube Quantum Resistors. *Science* **280**, 1744-1746 (1998).
19. I. van Weperen, S. R. Plissard, E. P. A. M. Bakkers, S. M. Frolov, L. P. Kouwenhoven, Quantized conductance in an InSb nanowire. *Nano Lett* **13**, 387-391 (2013).
20. G. Cheng *et al.*, Tunable Electron-Electron Interactions in $LaAlO_3/SrTiO_3$ Nanostructures. *Physical Review X* **6**, 041042 (2016).
21. J. F. Schooley, W. R. Hosler, M. L. Cohen, Superconductivity in semiconducting $SrTiO_3$. *Phys Rev Lett* **12**, 474-475 (1964).
22. G. Cheng *et al.*, Sketched oxide single-electron transistor. *Nat Nanotechnol* **6**, 343-347 (2011).

23. C. Cen *et al.*, Nanoscale control of an interfacial metal-insulator transition at room temperature. *Nature Materials* **7**, 298-302 (2008).
24. C. Cen, S. Thiel, J. Mannhart, J. Levy, Oxide nanoelectronics on demand. *Science* **323**, 1026-1030 (2009).
25. F. Bi *et al.*, "Water-cycle" mechanism for writing and erasing nanostructures at the $LaAlO_3/SrTiO_3$ interface. *Applied Physics Letters* **97**, 173110 (2010).
26. K. A. Brown *et al.*, Giant conductivity switching of $LaAlO_3/SrTiO_3$ heterointerfaces governed by surface protonation. *Nature Communications* **7**, 10681 (2016).
27. A. Annadi *et al.*, Quantized Ballistic Transport of Electrons and Electron Pairs in $LaAlO_3/SrTiO_3$ Nanowires. *Nano Lett* **18**, 4473-4481 (2018).
28. U. Schollwöck, The density-matrix renormalization group in the age of matrix product states. *Annals of Physics* **326**, 96-192 (2011).
29. U. Schollwck, The density-matrix renormalization group. *Reviews of Modern Physics* **77**, 259--315 (2005).
30. P. M. Ian, From density-matrix renormalization group to matrix product states. *Journal of Statistical Mechanics: Theory and Experiment* **2007**, P10014 (2007).
31. I. P. McCulloch, Infinite size density matrix renormalization group, revisited. *arXiv:0804.2509*, (2008).
32. G. M. Crosswhite, A. C. Doherty, G. Vidal, Applying matrix product operators to model systems with long-range interactions. *Phys Rev B* **78**, 035116 (2008).
33. J. A. Kjäll, M. P. Zaletel, R. S. K. Mong, J. H. Bardarson, F. Pollmann, Phase diagram of the anisotropic spin-2 XXZ model: Infinite-system density matrix renormalization group study. *Phys Rev B* **87**, 235106 (2013).
34. X.-W. Guan, M. T. Batchelor, C. Lee, Fermi gases in one dimension: From Bethe ansatz to experiments. *Reviews of Modern Physics* **85**, 1633-1691 (2013).
35. I. Affleck, Critical behaviour of SU(n) quantum chains and topological non-linear σ-models. *Nucl Phys B* **305**, 582-596 (1988).
36. Y. He, B. Tian, R. S. K. Mong, D. Pekker, *In preparation*.
37. P. W. Anderson, Model for electronic structure of amorphous semiconductors. *Phys Rev Lett* **34**, 953-955 (1975).
38. R. de-Picciotto *et al.*, Direct observation of a fractional charge. *Nature* **389**, 162 (1997).
39. L. P. Kouwenhoven, D. G. Austing, S. Tarucha, Few-electron quantum dots. *Rep Prog Phys* **64**, 701 (2001).
40. C. S. O'Hern, T. C. Lubensky, J. Toner, Sliding Phases in *XY* Models, Crystals, and Cationic Lipid-DNA Complexes. *Phys Rev Lett* **83**, 2745-2748 (1999).
41. D. Pekker, G. Refael, E. Demler, Finding the Elusive Sliding Phase in the Superfluid-Normal Phase Transition Smeared by *c*-Axis Disorder. *Phys Rev Lett* **105**, 085302 (2010).
42. P. Mohan, P. M. Goldbart, R. Narayanan, J. Toner, T. Vojta, Anomalously Elastic Intermediate Phase in Randomly Layered Superfluids, Superconductors, and Planar Magnets. *Phys Rev Lett* **105**, 085301 (2010).
43. C. L. Kane, A. Stern, B. I. Halperin, Pairing in Luttinger Liquids and Quantum Hall States. *Physical Review X* **7**, 031009 (2017).
44. E. Fradkin, *Field theories of condensed matter physics*. (Cambridge University Press, 2013).
45. U. Schollwöck, The density-matrix renormalization group. *Reviews of Modern Physics* **77**, 259-315 (2005).
46. P. Calabrese, J. Cardy, Entanglement entropy and quantum field theory. *Journal of Statistical Mechanics: Theory and Experiment* **2004**, P06002 (2004).

47. F. Pollmann, S. Mukerjee, A. M. Turner, J. E. Moore, Theory of Finite-Entanglement Scaling at One-Dimensional Quantum Critical Points. *Phys Rev Lett* **102**, 255701 (2009).

**Acknowledgements:**
This work is supported in part by a Vannevar Bush Faculty Fellowship ONR grant N00014-15-1-2847 (J.L.), Department of Energy Office of Science, Office of Basic Energy Sciences (DE-SC0014417, J.L.) and (DEFG02-06ER46327, C-B.E.), Air Force Office of Scientific Research (FA9550-15-1-0334, C.B.E.), and the Charles E. Kaufman Foundation (D.P).

# Supporting Information

## S1 Prior results on multi-fermion bound states

The stability of multi-fermion bound states depends critically on the spatial dimension. In higher spatial dimensions (*10*), three-component fermion systems with weak attractive interactions tend to form pairs and leave the remaining component unbound as opposed to forming trions (there is only one Fermi surface corresponding to the unbound component). Electron bound states formed by more than two electrons are much more stable in one dimension. The case of SU(*N*) symmetric, *N*-component fermionic systems in one dimension, with attractive interactions, has been extensively studied before. It was found that, unlike in higher dimensions, the *N*-component fermionic systems always form a Luttinger liquid consisting of *N*-fermion bound states (*34*). This phenomenon is among the most well understood ones of interacting quantum many-body physics, thanks to Bethe ansatz solutions (*10*) of certain models and conformal perturbation theory via either abelian (*44*) or non-abelian (*35*) bosonization.

SU(*N*) symmetry is generally absent in experiments, and therefore here we study the phase diagram of multicomponent systems in one dimension without this symmetry. We find that the trion phase is quite robust even in the absence of SU(3) symmetry. For example, it persists even when one of the interactions amongst the three types of particles is repulsive. This prediction is well supported by our density matrix renormalization group (DMRG) calculations (section S5) as well as our analysis of the low density case (section S6) of an extended three-component Hubbard model.

## S2 Effective 1D model of a waveguide with interacting electrons

In this section we derive an effective 1D interacting model suitable for DMRG analysis that we use to describe the phases near the sub-band bottom crossing points. Our analysis is not intended to be *ab-initio*. In constructing our model, we make a number of simplifying assumptions about the nature of the attractive electron-electron interaction, and therefore our numerical results should be seen as qualitative and not quantitative. Direct comparison to experimental data would require putting some complexity back into the model–like long-ranged interactions–as well as fitting of the model parameters. Rather, we aim to get the correct set of phases and the rough shape of the phase boundaries to justify our interpretation of the experimental data.

Our analysis begins with a derivation of a multi-band 1D continuous model starting from our 3D continuous Hamiltonian. We then introduce interactions and specify simplifying assumptions to ultimately arrive at an effective 1D lattice model. Importantly, the lattice spacing, and hence $U$ and $t$, in this model do not correspond to the actual microscopic values associated with LaAlO$_3$/SrTiO$_3$ but are rather effective parameters. We describe our DMRG analysis in the next section.

We start with Eq. (1) of the main text, in which we define the 3D continuous model of the electron waveguide without an interaction term. We reproduce it here for convenience:

$$H = \frac{(p_x - eB_z y)^2}{2m_x^*} + \frac{p_y^2}{2m_y^*} + \frac{p_z^2}{2m_z^*} + \frac{m_y^* \omega_y^2}{2} y^2 + \frac{m_z^* \omega_z^2}{2} z^2 - g\mu_B B_z s,$$
$$\text{where: } z > 0 \tag{S1}$$

Next, we solve this model to find the one-electron eigenstates (i.e. the waveguide modes):

$$\Phi_{\alpha, k_x}(x, y, z) = e^{ik_x x} \phi_{\alpha, k_x}(y, z), \tag{S2}$$

where, the index $\alpha = \{n_y, n_z, s\}$ denotes the waveguide mode and combines the transverse quantization index $n_y, n_z$ and the spin index $s$. We can then rewrite Hamiltonian (S1) as a one-dimensional, multi-band model:

$$H = \sum_{\alpha, k_x} \left( E_\alpha + \frac{\hbar^2 k_x^2}{2 m_x'} \right) c^+_{\alpha, k_x} c_{\alpha, k_x}, \tag{S3}$$

where $m_x' = m_x^* \Omega^2 / \omega_y^2$ is the effective mass, $\Omega = \sqrt{\omega_y^2 + \omega_c^2}$, $\omega_c = eB_z/\sqrt{m_x^* m_y^*}$, $E_\alpha$ is the sub-band bottom energy (defined in Eq. (2) of the main text), and $c^+_{\alpha, k_x}$ creates an electron in the single-particle state $|\Phi_{\alpha, k_x}\rangle$ defined in (S2). For completeness, we reproduce Eq. (2) of the main text

$$E_\alpha = \hbar\Omega \left( n_y + \frac{1}{2} \right) + \hbar\omega_z \left( (2n_z + 1) + \frac{1}{2} \right) - g\mu_B B_z s. \tag{S4}$$

We now introduce electron-electron interactions. As we do not know the microscopic origin, we begin with the most general form of the interactions

$$H_{\text{int}} = \sum_{\alpha,\beta,\gamma,\delta,k_1,k_2,k_3,k_4} \widetilde{U}_{\alpha,\beta,\gamma,\delta}(k_1, k_2, k_3, k_4)\, c^\dagger_{\alpha,k_1} c^\dagger_{\beta,k_2} c_{\gamma,k_3} c_{\delta,k_4}. \tag{S5}$$

At this point, we make some simplifying assumptions. First, we assume that the system is translationally invariant and hence $\widetilde{U}_{\alpha,\beta,\gamma,\delta}(k_1, k_2, k_3, k_4) \propto \delta(k_1 + k_2 - k_3 - k_4)$. Second, we ignore sub-band mixing $\widetilde{U}_{\alpha,\beta,\gamma,\delta}(k_1, k_2, k_3, k_4) \propto \delta_{\alpha,\delta} \delta_{\beta,\gamma}$. Finally, while $\widetilde{U}$ is generally a function of momentum, we assume that for the range of momenta that we are interested in $\widetilde{U}$ is momentum independent. Under these assumptions $H_{\text{int}}$ greatly simplifies

$$H_{\text{int}} = \sum_{\alpha,\beta,k_1,k_2,k_3} \widetilde{U}_{\alpha,\beta}\, c^\dagger_{\alpha,k_1} c^\dagger_{\beta,k_2} c_{\beta,k_3} c_{\alpha,k_1+k_2-k_3}. \tag{S6}$$

In order to analyze the model defined by equation (S3) and (S6) using DMRG we need to map onto an effective lattice model. Introducing an effective lattice length-scale $a$ (which is much larger than the underlying atomic lattice length-scale)--or equivalently momentum cut off $2\pi/a$--we obtain the lattice Hamiltonian

$$H = \sum_{\alpha,j}(-\mu + E_\alpha - 2t_\alpha)c^\dagger_{\alpha,j}c_{\alpha,j} - \sum_{\alpha,j}\left(t_\alpha c^\dagger_{\alpha,j}c_{\alpha,j+1} + h.c.\right) + \sum_{\alpha,\beta,i} U_{\alpha,\beta}c^\dagger_{\alpha,i}c^\dagger_{\beta,i}c_{\beta,i}c_{\alpha,i}, \quad (S7)$$

where the first term describes the effective chemical potential for sub-band $\alpha$, the second term describes the low momentum dispersion of the electrons in sub-band $\alpha$ (i.e., $t_\alpha = \frac{\hbar^2}{m'_\alpha a^2}$), and the final term is the Fourier transform of Eq. (S6) that describes the effective electron-electron interactions.

We also assume that the mass and interaction $t_\alpha$ and $U_{\alpha,\beta}$ are weakly varying with the magnetic-field and chemical potential in the narrow region of interest and hence we assume that they are constant ($B$ in the vicinity of $B_{Pa}$, and $\mu \ll$ energy cut off). In our numerics we introduce $\mu_\alpha = \mu - E_\alpha + B(\partial_B E_\alpha)_{B \to B_0}$ as the combination of the chemical potential and the magnetic energy composed of a Zeeman and an orbital part. We performed zero-temperature DMRG (*5, 28, 30-33, 45*) calculations for the above Hamiltonian (S1) with fixed electron filling fractions for each spin and subband. We identify the phases by measuring the central charge (number of gapless degrees of freedom) (*46, 47*) and also the correlation functions. An artifact of this lattice model, not present in experiment, is that at large chemical-potential components can become completely filled which causes them to gap out.

### S3 Model parameters for iDMRG
We perform infinite Density Matrix Renormalization Group (iDMRG) calculations for fixed filling fractions in the unit of 1/41 for both 2-subband and 3-subband case. For the 2-subband case (Figure 5A) we set: $t_1 = t_2 = 1$, $U_{1,2} = -1$ and $g_1 = 1, g_2 = -1$. For the 3-subband case (Figure 5B) we set: $t_1 = t_2 = t_3 = 1$, $U_{1,2} = U_{1,3} = U_{2,3} = -1$ and $g_1 = 0.1, g_2 = 0.4, g_3 = 0.5$. The bond dimensions we use are 14, 20, 28, 40, 57, 80, 113.

### S4 DMRG simulation details and methods
In this section we present the details for analyzing the ground state properties in our DMRG calculation. We first extract the ground state energy using finite-size scaling; then we analyze equal time correlation functions and central charge to determine the phase of the ground state.

We extract the ground state energy by performing finite bond dimension scaling: specifically, we fit the energy as a function of the bond-dimension dependent correlation length $\xi$ (*47*), as shown in Figure S1A. The error bar on the extracted $E_\infty$, with 95% certainty, is in the order of $10^{-6}t$.

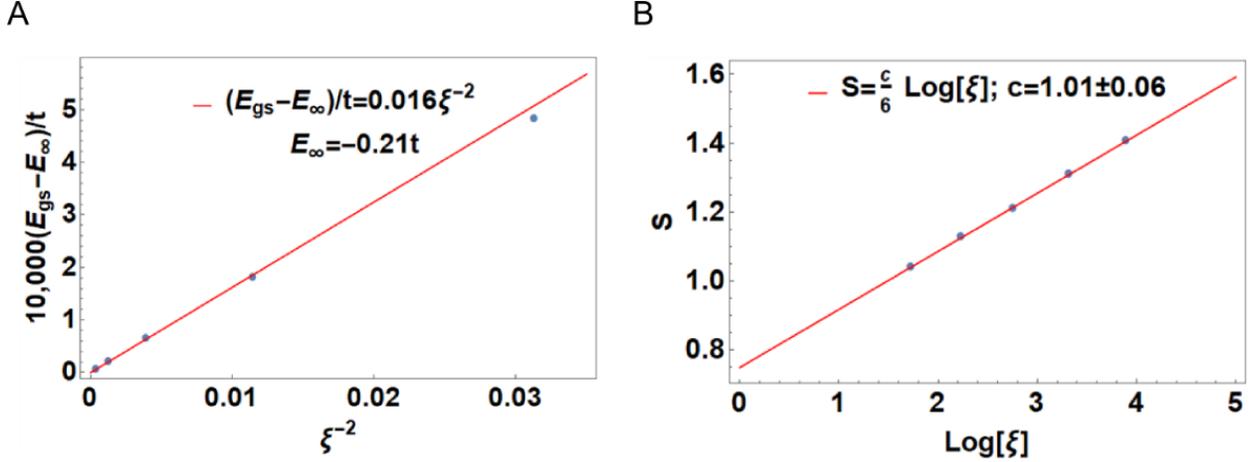

Figure S1 **Finite bond-dimension scaling in iDMRG analysis of the pair phase.** (A) Ground state energy vs the DMRG correlation length. Each point corresponds to a different DMRG bond dimension. The true ground state energy $E_\infty$ is extrapolated from $\xi \to \infty$ using the fact that the energy operator has scaling dimension 2. (B) Bipartite entanglement entropy vs the logarithm of the DMRG correlation length. Again, each point corresponds to a different DMRG bond dimension. The slope gives the central charge see *(46, 47)*. (parameters used: $t_1 = t_2 = 1$, $U_{1,2} = -2$, $n_1 = n_2 = 7/29$, bond dimension=*28, 40, 57, 80, 113* for the five points).

To analyze the ground state properties, we first determine the behavior of the equal time correlation functions; this is most convenient for iDMRG. In the language of Luttinger liquid theory, the formation of $q-$electron bound states corresponds to the attractive interaction gapping out "spin" excitations *(42)* and the gapless excitation (operator) must involve $q$ particles from different subbands together. This is to say that whether an operator is gapless/gapped can be determined by the algebraic/exponential decay behaviors of its two-points correlation functions. For example, for a paired state, the single-electron correlation function $\langle \psi_\sigma^\dagger(r)\psi_\sigma(0)\rangle$ decays exponentially in $r$, so the single-electron excitations are gapped. The pair correlation $\langle \psi_\alpha^\dagger(r)\psi_\beta^\dagger(r)\psi_\beta(0)\psi_\alpha(0)\rangle$, however, decays algebraically, indicating that there is no energy gap to adding an electron pair. Similarly, in the trion phase, the single- and pair-correlator both decays exponentially, while the three-electron correlators are algebraic. Our iDMRG data indeed shows this behavior, as an example we plot the one-, two-, and three-particle correlation functions in the trion phase (Figure S2) and observe that while the first one- and two-particle correlation functions decay exponentially the three-particle correlation function decays algebraically.

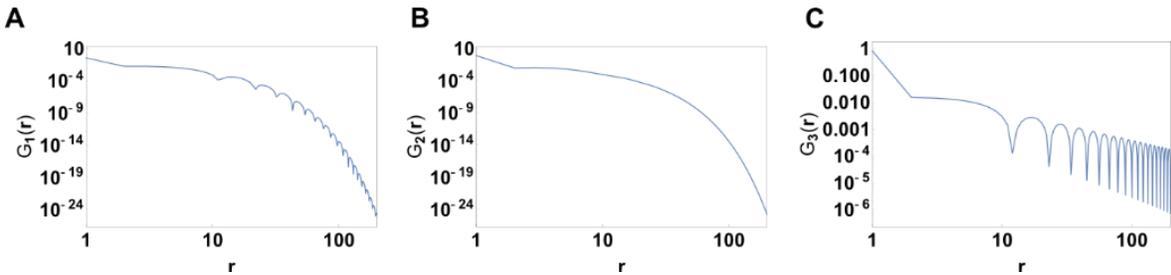

Figure S2 **Correlation functions of a quantum ground state in trion phase.** The plots show that single-electron correlation functions (A) as well as pair correlation function (B) decay exponentially while the trion correlation function (C) decays algebraically (parameters used: $t_1 = t_2 = t_3 = 1$, $U_{1,2} = U_{1,3} = U_{2,3} = -1$, $n_1 = n_2 = n_3 = 1/11$, bond dimension= 800).

Besides the correlation functions the other indicator of the quantum state is the central charge $c$. For the phases relevant to this work, $c$ counts the total number of species of gapless particles. To compute the central charge, we do a linear fit of the entanglement entropy $S$ v.s. $\text{Log}(\xi)$ [where $\xi$ is the bond dimension dependent correlation length]. As an example, the central charge fitting for a pair phase is shown in Figure S1B where the central charge is 1. For the phases shown in Figure 5, we find that F, P, and T phases have central charge 1; 2F, P+F have central charge 2; 3F has central charge 3.

The liquid phases can also be diagnosed by looking for kinks in the energy per site $E(n_1, n_2, n_3)$ as a function of particle densities ($n_1$, $n_2$, and $n_3$). To obtain the energy as a function of the chemical potentials ($\mu_\alpha$'s of Eq. (S7)), we introduce the function $G$

$$G(\{\mu_\alpha\}, \{n_\alpha\}) = E(\{n_\alpha\}) - \sum_\alpha \mu_\alpha n_\alpha \tag{S8}$$

and perform a Legendre transform

$$\Phi(\{\mu_\alpha\}) = \min_{n_\alpha} G(\{\mu_\alpha\}, \{n_\alpha\}). \tag{S9}$$

In our numerics $n_\alpha$'s are integer multiples of 1/41, where 41 is our iDMRG unit cell size. In Figure S3A we plot $G(\{\mu_\alpha\}, \{n_\alpha\})$ as a function of $n_3$ for fixed $n_1$, $n_2$ and a particular choice of $\mu_\alpha$'s. We observe that there is a kink in $G$ at the point $n_1 = n_2 = n_3$. This kink is indicative of an extended trion phase. The single particle-excitation gap is given by the left derivative (hole-like excitation) and right derivative (particle-like excitation) of $G$ at the kink. In Figure S3B we plot the minimum of the 1 and 2 particle gaps for a line cut across the trion phase. Showing that 1 and 2 particle excitation are gapped throughout this phase.

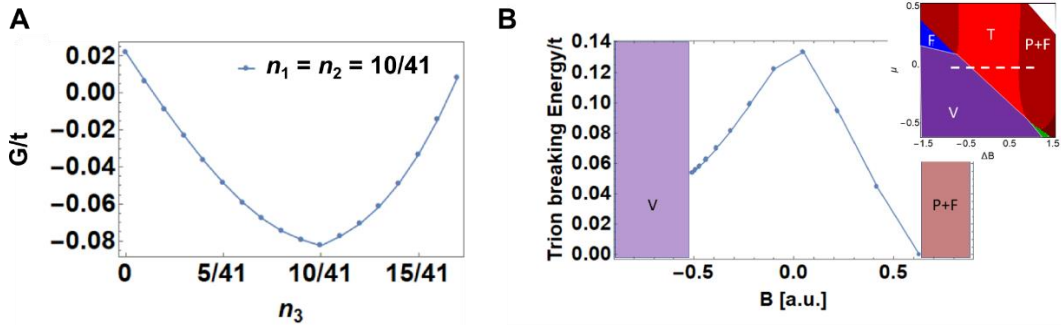

Figure S3 (**A**) DMRG ground state energy as a function of filling $n_3$ with $n_1$ and $n_2$ fixed as indicated. The minimum of the ground state energy at $n_3 = n_2 = n_1$, along with the kink in the energy at the same point, indicates that the trion phase is stable. (**B**) Trion phase is characterized by a finite one and two particle gaps. In this figure we plot the minimum energy to insert/remove one or two particles as a function of the magnetic field. The plot indicates that along this particular cut through the phase diagram the trion phase is stable for $-0.5 < B < 0.6$ as the one and two particle gaps are always finite. The line cut is made along $\mu = 0$, where on the very left side is the vacuum phase and on the right side is the P+F phase (see the inset for the location of the cut in which we reproduce a portion of the phase diagram from the main text Fig. 5). (Parameters used to construct this figure are listed in section S3)

## S5 Extended DMRG data set

Here we present additional DMRG phase diagrams which explicitly demonstrates that interactions can stabilize a $3e^2/h$ conductance step. In the non-interacting model, the presence of the $3e^2/h$ conductance step relies on three sub-band bottoms (SBBs) intersecting at a single point in the chemical potential-magnetic field plane. However, this intersection of three lines at a single point in the plane is not generic. We now show that the $3e^2/h$ conductance step can persist in the interacting model even when we shift the sub-bands so that all three SBBs do not intersect at a single point. Without loss of generality, we consider the model (equation S1) with $t_1 = t_2 = t_3 = 1$, $U_{1,2} = U_{1,3} = U_{2,3} = U$, $\mu_1 = \mu - B - \delta$, $\mu_2 = \mu$, and $\mu_3 = \mu + B - \delta$. Here, as before B represents the magnetic field and $\mu$ the chemical potential. The parameter $\delta$ adjusts how close the three SBBs come to intersecting at a single point in the chemical potential-magnetic field plane.

As a starting point, consider the case $\delta = 0$. In this case, in the absence of interactions, the three SBBs intersect at the point $\mu = B = 0$ (Figure S4B). Turning on interactions, by setting $U = -1$, results in the phase diagram depicted in Figure S4E. We observe that the trion phase, and hence the $3e^2/h$ conductance jump, is indeed stable over a finite range of magnetic fields. Let us now consider the case $\delta = -0.2$. In the absence of interactions, the three SBBs no longer intersect at a single point (Figure S4A) and hence the $3e^2/h$ conductance jump disappears. Turning on attractive interactions the trion phase and the $3e^2/h$ conductance jump reemerge, although over a narrower range of B (Figure S4D). Similarly, setting $\delta = 0.2$ we observe that the three SBBs no longer intersect at a single point in the absence of interactions (Figure S4C), but the trion phase and the $3e^2/h$ conductance jump returns once attractive interactions are turned on (Figure S4F).

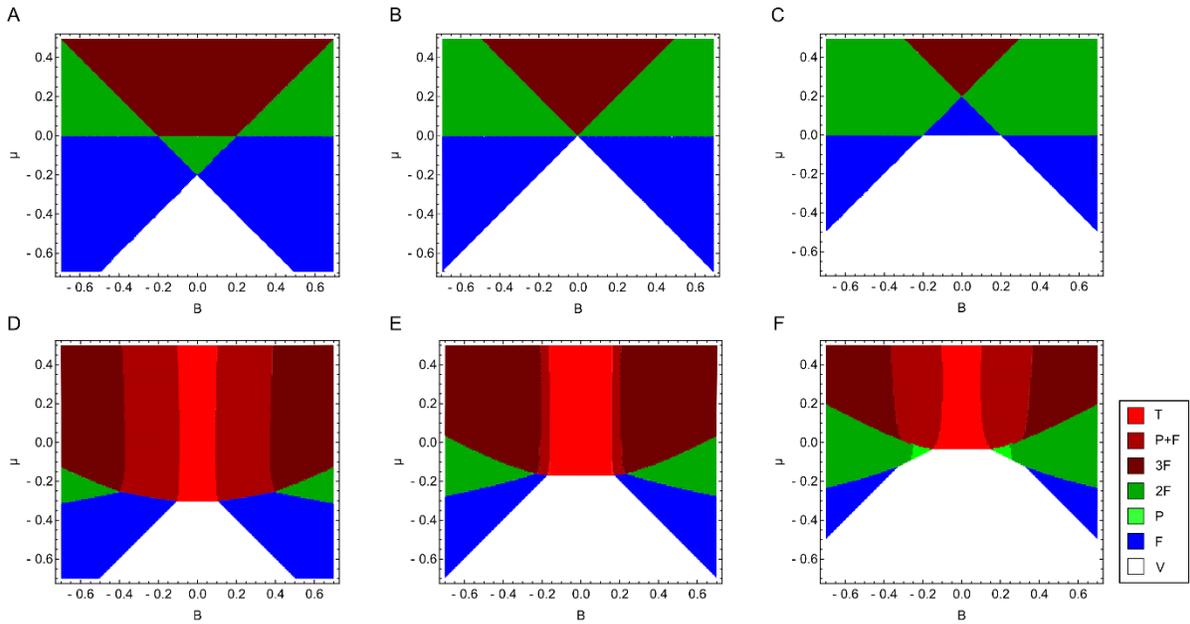

Figure S4 **Extended DMRG data.** (**A-C**) DMRG phase diagrams for a non-interacting model. (A) $\delta = -0.2$, (B) $\delta = 0$, and (C) $\delta = 0.2$, for non-zero values of $\delta$ the sub-bands are shifted so that all three SBBs do not intersect at a single point. (**D-F**)

DMRG phase diagrams with the attractive interactions turned on. (D) $\delta = -0.2$, (E) $\delta = 0$, and (F) $\delta = 0.2$. For all cases with attractive interactions there is the formation of a stable trion phase over a range of magnetic field values. Abbreviations for various phases: $m$F: $m$ distinct fermi surfaces, P: paired phase, T: trion phase, V: vacuum, A+B: phase composed of A and B. (Parameters used to construct this figure are listed in section S5, finite bond dimension scaling was done using bond dimensions listed in section S3)

## S6 Trions at low filling and weak interactions

For weak attractive interactions, the magnetic field scale over which locking occurs is set by the pair, trion, quaternion, ... binding energy scale $U^2/t$, where $U < 0$ is the strength of the local attractive interaction and $t$ is the hopping matrix element. We now present a more detailed answer.

The phenomenon of sub-band bottom locking originates in few-particle physics, as it occurs at very low filling. To illustrate how the magnetic field scale over which locking occurs depends on both the attractive interaction strength and the degeneracy, we calculate the binding energy of a single pair and a single trion. For the case of pairs, the locking scale [the magnetic-field scale at which the pair phase gives way to the 1F phase] can be found by comparing the energy of the pair to the energy of a single particle

$$E_{\text{pair}} + (g_1 + g_2)(B - B_0) = \min\{g_1, g_2\}(B - B_0), \qquad (S10)$$

where $E_{\text{pair}} < 0$ is the pair binding energy, $B_0$ is the magnetic field at which the two sub-band bottoms cross, and $g_i = d\epsilon_i/dB$ is the effective g-factor for sub band $i$. For the case of trions, we must compare the energy of the trion to the energy of a single particle and of a pair

$$\begin{aligned} E_{\text{trion}} + (g_1 + g_2 + g_3)(B - B_0) = \min\{ &g_1(B - B_0), g_2(B - B_0), g_3(B - B_0), \\ &E_{\text{pair 1,2}} + (g_1 + g_2)(B - B_0), \\ &E_{\text{pair 2,3}} + (g_2 + g_3)(B - B_0), \\ &E_{\text{pair 1,3}} + (g_1 + g_3)(B - B_0)\}, \end{aligned} \qquad (S11)$$

We now compute $E_{\text{pair}}$ for two dissimilar fermions. For the case of the Hubbard model, the most efficient way to find the binding energy is to look for zeros of the T-matrix. The equation for the zeros (at zero center of mass momentum) is

$$U^{-1} + \int_{-\pi}^{\pi} \frac{dk}{2\pi} \frac{1}{2\epsilon_k - E_{\text{pair}}} = 0, \qquad (S12)$$

where $\epsilon_k = 2t(1 - \cos(k))$ is the kinetic energy of a fermion with momentum $k$. Solving this equation, we find

$$E_{\text{pair}} = 4t - \sqrt{16t^2 + U^2}. \qquad (S13)$$

For weak attractive interactions $E_{\text{pair}} \approx -U^2/8t$, which gives the magnetic field scale.

We are not aware of an analytical solution for $E_{trion}$, and hence resort to computing it numerically. In our model, the particles interact pair-wise, and thus the trion binding energy depends on the three interaction strengths $U_{12}$, $U_{23}$, and $U_{13}$. In Figure S5 we compare the trion and pair binding energy for several values of interaction parameters (as we do not know these *ab initio*). We observe that in all cases, for weak interactions the binding energy scales with $U^2/t$. For the case of weak symmetric attractive interactions ($\{U_{12}, U_{23}, U_{13}\} = \{U, U, U\}$) the trion binding energy is approximately four times stronger than the pair binding energy. Interestingly, we find that even if two components of the trion repel each other ($\{U_{12}, U_{23}, U_{13}\} = \{U, -U, U\}$), the trion still has a lower binding energy than the pair, indicating that there is a stable trion phase.

When $U_{12} \sim U_{23} \sim U_{13} \sim U_{pair}$ the magnetic field scale over which the trion phase remains locked is larger than the locking scale for the pair phase. Experimental observations are consistent with these findings (see Figure 4B,C).

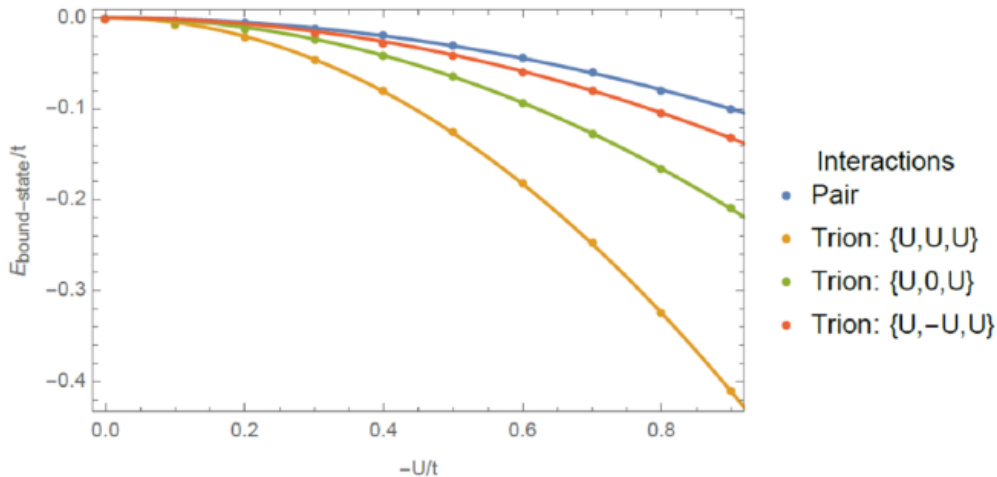

Figure S5 **Comparison of Pair and Trion binding energies as a function of interaction strength**. For the case of trions, we consider three cases: $\{U_{12}, U_{23}, U_{13}\} = \{U, U, U\}$, $\{U, 0, U\}$, and $\{U, -U, U\}$ as labeled. The trion binding energies were computed on a 55-site Hubbard model lattice with periodic boundary conditions.

**S7 Anharmonic corrections to our single-particle model**
In this section, we consider perturbations of the single-particle Hamiltonian in Eq. (1) of the main text and Eq. (S1) of the supplement. In the unperturbed single-particle model, the SBB crossings which form the Pascal liquid states all appear at a fixed magnetic field, $B_{Pa}$ (Figure S6A) while, experimentally, $B_{Pa}$ may shift as a function of $\mu$ (Figure 2 of the main text). By including anharmonic terms in the single-particle model, it is possible to achieve this effect.

We begin by introducing a quartic term to the single-particle Hamiltonian, giving

$$H = \frac{(p_x - eB_z y)^2}{2m_x^*} + \frac{p_y^2}{2m_y^*} + \frac{p_z^2}{2m_z^*} + \frac{m_y^* \omega_y^2}{2} y^2 + w_4 y^4 + \frac{m_z^* \omega_z^2}{2} z^2 - g\mu_B B_z s, \quad (S14)$$

where $w_4$ is the strength of the quartic confinement. We find the SBBs defined by Eq. (S14) by numerically solving for the transverse waveguide modes that correspond to $k_x = 0$. The resulting SBB energies are shown in Figure S6. With the addition of $w_4$, the Pascal series is lost due to the shifting of the SBBs (Figure S6B). As the model parameters are tuned, however, the Pascal series is approximately recovered and $B_{Pa}$ begins to shift with $\mu$ as observed in experiment (Figure S6C,D).

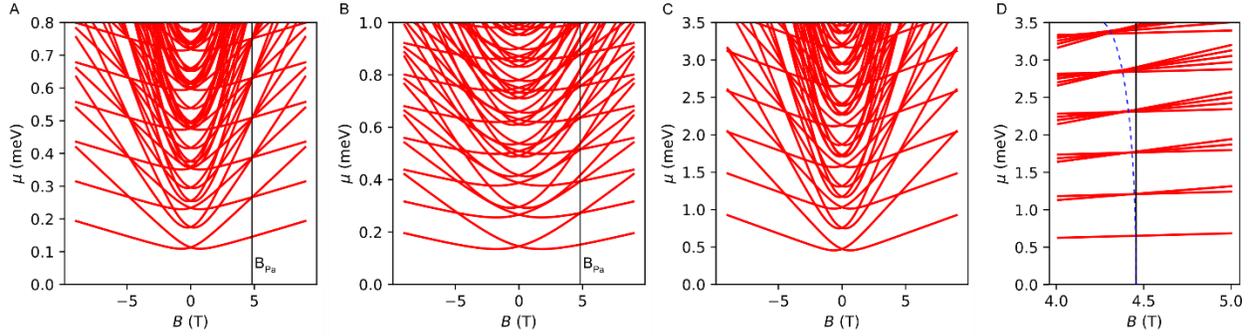

Figure S6 **Single-particle model with an anharmonic term**. (**A**) The SBB energies when $m_x = m_y = 2.34 m_e$, $\omega_y = 81$ GHz, $\omega_z = 96$ GHz, and $g = 0.44$. This set of parameters is tuned to the Pascal Condition, $\Omega(B_{Pa}) = 4\omega_z = 2g\mu_b B_{Pa}$, leading to the multiple SBB crossings at $B = B_{Pa}$. (**B**) The SBB energies when an anharmonic term $w_4 y^4$ with $w_4 = 5.45 \times 10^{29}$ meV/$m^4$ is added. The shift destroys the SBB crossings at $B = B_{Pa}$ as shown by the line. (**C**) The SBB energies when $m_y = 0.46 m_e$, $\omega_y = 81$ GHz, $\omega_z = 470$ GHz, $g = 2.17$, and $w_4 = 1.02 \times 10^{29}$ meV/$m^4$. With this set of parameters, the Pascal series is approximately realized, but now with a Pascal field $B_{Pa}$ which is a function of the SBB crossing instead of constant as in the harmonic case, following the blue dashed line rather than the black line (previous $B_{Pa}$ value) (**D**).

### S8 Fits of transconductance data (Figure 4)

To find the strength of the locking for the pair and trion phases we performed a nine parameter non-linear least squares fit of the transconductance data, where one of the parameters in the model is the magnitude of the locked region. The nine parameters are used to create a profile of the phase, where two (pair) or three (trion) states come together and are locked over a range of magnetic field values. The nine parameters include the critical point at the center of the locked region, $(B_c, \mu_c)$, the magnitude and orientation of the locked region, the locations of each state at $B = 0$ T, and the magnitude and width of a $\sec^2(x)$ function, which is how the transconductance data is modeled. The profile used for fitting a trion state can be seen in Figure S7. The fit was performed by minimizing the error function, which is the sum of the square of the difference between the transconductance data and the profile. To calculate the uncertainty of the results of the fit we calculated the Hessian of the error function.

Results for the magnitude of the locked state are quoted in the main text, and plots of the fit for the pair and trion states of Device F are plotted in the main text in Figure 4.

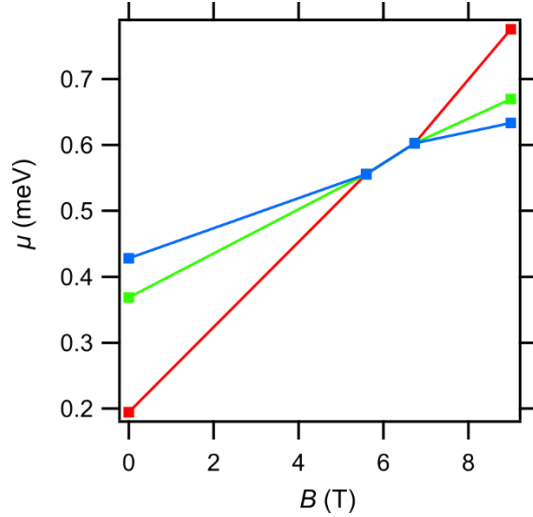

Figure S7 **Model for transconductance fitting**. Profile created from the nine parameter model to fit the transconductance data for the trion state. The results of the fit show a finite magnitude for the locked region.

**S9 Fits of conductance data (Figure 6)**

We also fit the experimental data using a second method. The second method was used to fit the angle dependent data shown in Figure 6, and find the strength of the reentrant pairing for the $|0, 0, \downarrow\rangle$ and $|0,1, \uparrow\rangle$ subbands. The second fitting method involves a line by line nonlinear least squares fit of the conductance data where each step in the conductance is modeled as $\tanh(x)$ function, so that the total conductance is given as:

$$G = \frac{A_1}{2}\left(1 + \tanh\left(\frac{x - v_1}{w_1}\right)\right) + \frac{A_2}{2}\left(1 + \tanh\left(\frac{x - v_2}{w_2}\right)\right) + \cdots \quad (S17)$$

where $G$ is the total conductance, $A_i$ is the amplitude of the $i^{th}$ step, $v_i$ is the position of the step, and $w_i$ is the width of the step. Typical conductance data is shown in Figure S8. The step amplitude is restricted to be within 0.9 and 1.1 $e^2/h$. To determine how many steps are needed for each curve, the fit is performed with a range of possible steps, based on the maximum conductance of the curve, and the fit with the smallest residue (weighted mean square error between the best fit and the data) is chosen. The conductance curve at each magnetic field value was fit independently and by taking the difference between the peak positions of the second and third steps (the subbands that for the first paired state) we are able to determine the magnetic field value where the states pair (where the step positions are the same) and where the states break apart again (shown with red and blue circles in Figure 6D).

The transconductance model fitting method was also used to fit the angle dependent data in Figure 6 as a comparison of the two fitting methods. This method found a pairing magnitude of $0.63 \pm 0.29$ T, $0.41 \pm 0.07$ T, and $0.68 \pm 0.08$ T for $\theta = 0°, \; 10°,$ and $20°$. Which agrees with the pairing strength found with the line by line conductance fitting method.

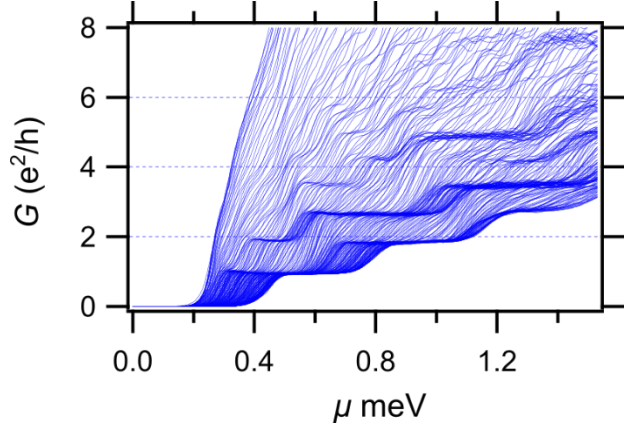

Figure S8 **Conductance curves**. Typical conductance curves at different out-of-plane magnetic field values. Fits of these curves, from Device G, were used to find the locations of each step and the strength of the locking of the subbands for the paired states of the angle dependent data in Figure 6.

## S10 Additional rotation data

Additional rotation data for device G can be found in Figure S9. Figure S9A shows a transconductance map of for device G at an angle of $\theta = 10°$, $\phi = 0°$ which corresponds to rotating in the XZ plane, where the in-plane component of the B field is parallel to the waveguide. This can be compared to Figure 6D $\theta = 10°$, where the rotation is the YZ direction. The two transconductance maps look essentially identical. Figure S9B shows a sweep of the in-plane angle from $\phi = -90°$ to $\phi = 90°$, at a fixed field of $B = 18$ T. Figure S9C shows the transconductance map for $\theta = 90°$ and $\phi = 90°$ so the magnetic field is in-plane, perpendicular to the waveguide. As the out-of-plane magnetic field component is decreased, the SBBs become more closely spaced, and are hard to distinguish.

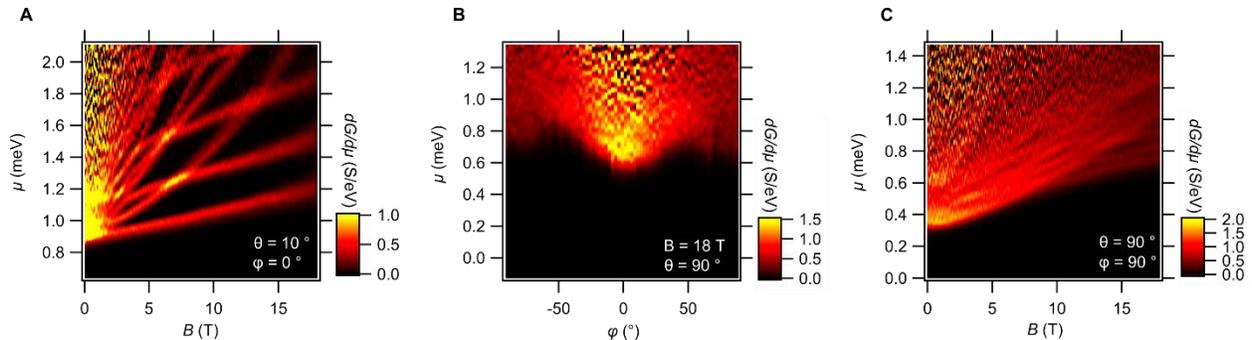

Figure S9 **Additional rotation data.** (**A**) Transconductance map for device G for angle $\theta = 10°$ and $\phi = 0°$, corresponding to a rotation in the XZ plane. (**B**) Rotation of the in-plane angle $\phi$ from $-90°$ to $90°$ at $\theta = 90°$ (in-plane) and a fixed magnetic field $B = 18$ T. (**C**) Transconductance map for device G for angle $\theta = 90°$ and $\phi = 90°$ (B is oriented in-plane, perpendicular to the waveguide). As the angle is rotated into the plane of the sample, the subbands become more closely spaced, with a completely in-plane B field orientation they are virtually indistinguishable.

## S11 Additional line cuts

Figure S10 shows additional conductance data for devices A-G highlighting the Pascal series of conductance steps for each device. Each plot shows two line cuts, taken at a constant magnetic field (red) and a cut where the magnetic field is varied so that the line cut is angled to pass through the Pascal degeneracies which vary slightly with magnetic field for each sample.

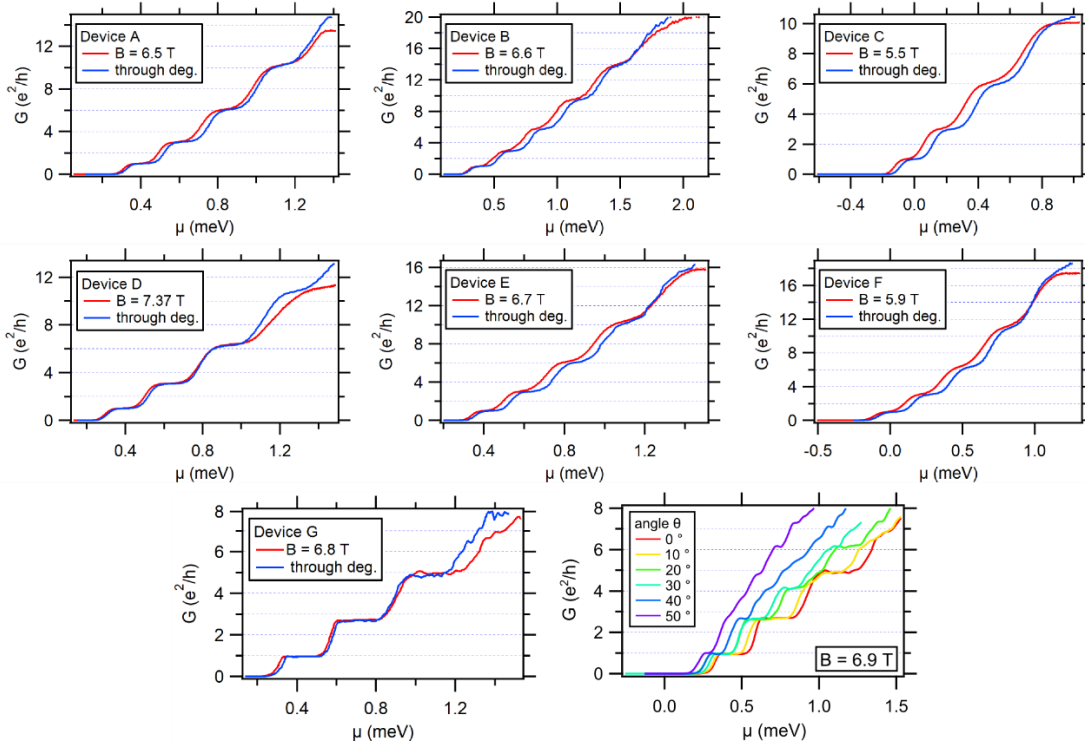

Figure S10 **Additional line cuts for Devices A-G**. Panels for device A-G include a line cut at a constant magnetic field (red) and a line cut where the magnetic field is varied so that the line cut passes through each Pascal degeneracy (blue). The last panel includes line cuts from device G at a constant magnetic field magnitude $B = 6.9$ T for an increasing angle $\theta$.

## S12 Width of transconductance peaks

Due to the intensity plot representation of the transconductance, it may appear that the width of the transconductance peaks increases around Pascal state where multiple SBBs converge and lock together. Figure S11 shows the transconductance intensity map for simulated single-particle model data where the Pascal condition has been satisfied (as in Figure 3A). Here it also appears that the width of the SSBs increase near the crossing point, despite the fact that all the individual subbands have a fixed width, and are all crossing at a single point, the Pascal field. We have also included a transconductance line cut (Figure S12A) from device A, at $B = 6.5$ T, corresponding to the conductance line cut in Figure 1A. Each peak is fit with a Lorentzian function, and the normalized fits are plotted in Figure S12B. The widths of the normalized peaks for the first three degeneracies are similar.

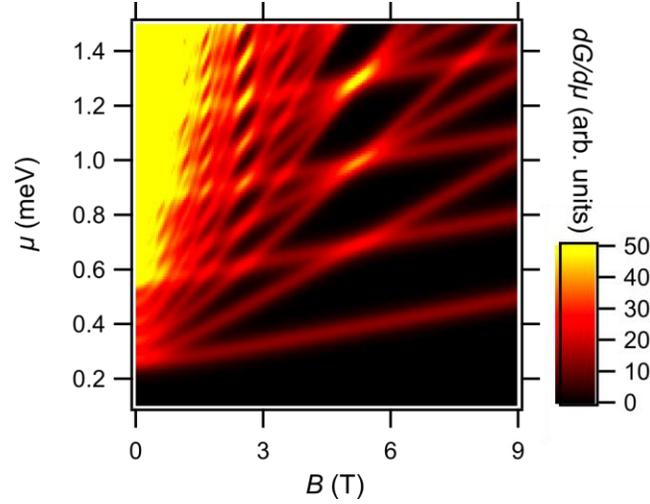

Figure S11 **Single particle model transconductance intensity plot.** Plot of the single particle model energy bands as an intensity plot. Here the Pascal condition has been satisfied, as in Figure 3A, so that the subbands all cross at the Pascal field $B_{Pa}$. Due to the nature of the intensity plot, where the subbands cross it appears that the width of the lines in increasing, despite the fact that the width of each line is a fixed value, and they all cross at the Pascal field.

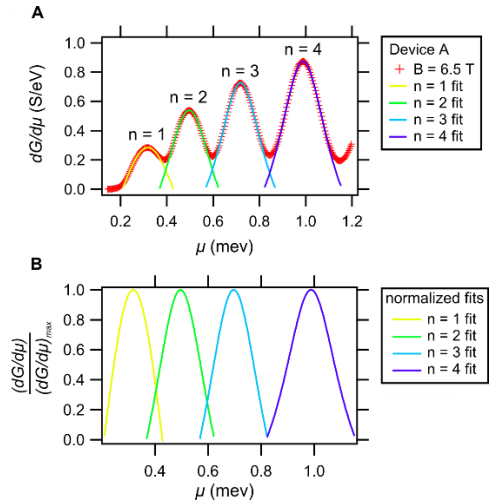

Figure S12 **Representative transconductance curve.** (A) Transconductance line cut for device A at $B = 6.5$ T, corresponding to the conductance line cut in Figure 1A. Fits for each individual peak to a Lorentzian function. (B) Normalized fits for each degeneracy. The first three fits for $n = 1, n = 2,$ and $n = 3$ are of similar widths. The width fit for $n = 4$ is slightly larger.

**Supplemental movie S1:** https://youtu.be/yBz3HaNi4jE
This movie shows the transconductance data of device G as the applied magnetic field of 18 T is rotated in the YZ plane from an out-of-plane to an in-plane orientation.